\documentclass[prb, aps,amsmath,amssymb,amsfonts, twocolumn]{revtex4}
\usepackage{graphicx}
\usepackage{bm}
\usepackage{amssymb}
\usepackage{amsmath}
\usepackage{amsfonts}

\newcommand {\bkt} [1] {\langle #1 \rangle}

\newcommand {\pd} [2] {\frac{\partial #1}{\partial #2}}

\DeclareMathOperator{\trace}{tr}

\begin{document}
\title{Spin polarization decay in spin-1/2 and spin-3/2 systems}
\author{Dimitrie Culcer}
\author{R. Winkler}
\affiliation{Advanced Photon Source, Argonne National Laboratory,
Argonne, IL 60439} \affiliation{Department of Physics, Northern
Illinois University, DeKalb, IL 60115}
\begin{abstract}
  We present a general unifying theory for spin polarization decay
  due to the interplay of spin precession and momentum scattering
  that is applicable to both spin-1/2 electrons and spin-3/2 holes.
  Our theory allows us to identify and characterize a wide range of
  qualitatively different regimes. For strong momentum scattering or
  slow spin precession we recover the D'yakonov-Perel result,
  according to which the spin relaxation time is inversely
  proportional to the momentum relaxation time. On the other hand,
  we find that, in the ballistic regime the carrier spin
  polarization shows a very different qualitative behavior. In
  systems with isotropic spin splitting the spin polarization can
  oscillate indefinitely, while in systems with anisotropic spin
  splitting the spin polarization is reduced by spin dephasing,
  which is non-exponential and may result in an incomplete decay of
  the spin polarization. For weak momentum scattering or fast spin
  precession, the oscillations or non-exponential spin dephasing are
  modulated by an exponential envelope proportional to the momentum
  relaxation time. Nevertheless, even in this case in certain
  systems a fraction of the spin polarization may survive at long
  times. Finally it is shown that, despite the qualitatively
  different nature of spin precession in the valence band, spin
  polarization decay in spin-3/2 hole systems has many similarities
  to its counterpart in spin-1/2 electron systems.
\end{abstract}
\date{\today}
\maketitle

\section{Introduction}

The achievement of a lasting spin polarization has been a
long-standing goal in semiconductor physics. Successful efforts to
\emph{generate} a spin polarization magnetically, optically and
electrically, have yielded a steady stream of novel physics and
promising applications. \cite{zut04, wol01} Ferromagnetic
semiconductors are edging towards room temperature \cite{mac06} and
spin currents have been measured directly. \cite{tin06} Successes
such as these have turned semiconductor spin electronics into a
vibrant and rewarding area of research, as well as a promising
candidate for novel information processing methods.

Both for fundamental physics and for technological applications, it
is important to know how to \emph{maintain} a spin polarization once
it is generated. Therefore, a detailed understanding of the
mechanisms leading to spin polarization decay is critical in all
areas mentioned above. In the return to equilibrium of an excess
spin polarization spin-orbit interactions play an important role.
Spin-orbit coupling always gives rise to spin precession, and the
interplay of spin precession and momentum scattering is frequently
the main cause of spin polarization decay. \cite{dya72, dya86,
pik84} A spin polarization in a semiconductor may also decay via
spin flips induced by momentum scattering or by exchange
interactions, though these mechanisms have a more limited range of
applicability. \cite{pik84, ave02, ell54, yaf63}

Spin relaxation in spin-1/2 electron systems has received
considerable attention. \cite{dya72, dya86, pik84, dya71, dya06,
kis00, song02, hua03, ohn99, sli90, dzh04, ave99, ave02, ave06,
gri02, ell54, yaf63, qi03, mis04, lau01, kai04, bro04, ble04,
ohno07} For electrons the spin-orbit interaction can always be
represented by a Zeeman-like Hamiltonian $H = (\hbar/2) \,
\bm{\sigma}\cdot\bm{\Omega} (\bm{k})$ describing the interaction of
the spin with an effective wave vector-dependent magnetic field
$\bm{\Omega} (\bm{k})$. The electron spin precesses about this field
with frequency $\Omega \equiv |\bm{\Omega} (\bm{k})|$. An important
parameter is the product of the frequency $\Omega$ times the
momentum relaxation time $\tau_p$. In the ballistic (clean) regime
no scattering occurs and the temperature tends to absolute zero, so
that $\Omega\tau_p \rightarrow \infty$. The weak scattering regime
is characterized by fast spin precession and little momentum
scattering due to, e.g., a slight increase in temperature, yielding
$\Omega \, \tau_p \gg 1$. In the strong momentum scattering regime
$\Omega \, \tau_p \ll 1$.

Electron systems are often in the strong scattering regime. In this
case the main mechanism leading to spin polarization decay is the
D'yakonov-Perel (DP) mechanism, \cite{dya72, dya86, pik84} which was
shown to be dominant over a wide range of temperatures \cite{pik84}
and, for particular forms of $\bm{\Omega} (\bm{k})$, to lead to a
noticeable anisotropy in the relaxation times for different spin
components \cite{ave99, ave06} and anisotropic spin diffusion.
\cite{gri02} Most past work has concentrated on this regime. On the
other hand, in recent years state-of-the-art technology has enabled
the growth of ballistic samples which have been at the forefront of
spin-related experiments. \cite{zut04, wol01} Yet spin polarization
decay in ballistic spin-1/2 systems has received comparably little
attention \cite{bro04, ble04, zut04} and has been considered
recently mostly in the context of spin transport in an electric
field. \cite{qi03, mis04}

For spin-3/2 holes the spin-orbit interaction cannot be written as
an effective field, and spin precession is qualitatively different.
\cite{dim06} Since spin-orbit coupling is more important in the
valence band, hole spin information is lost faster, and the relative
strengths of spin-orbit coupling and momentum scattering can vary.
Yet spin relaxation of spin-3/2 holes has also been studied to a
lesser extent, both experimentally \cite{hil02} and theoretically.
\cite{ave02, dya84a, ser05, lu06, uen90, fer91} A theory of spin
relaxation valid for electrons and holes in all regimes of momentum
scattering does not, to our knowledge, exist to date.

With these observations in mind, we present in this article a
general unifying quantitative theory for the return to equilibrium
of excess spin polarizations in the conduction and valence bands of
semiconductors brought about by the interplay of spin precession and
momentum scattering. We do not rely on the assumption, made in most
previous work, \cite{dya72, dya86, pik84, ave99, dya71, dya06,
ave02, ave06, qi03, lau01, kai04} that $\Omega \, \tau_p \ll 1$. We
demonstrate that spin polarization decay in different regimes of
momentum scattering in spin-1/2 electron and spin-3/2 hole systems
contains considerable rich and novel physics. For example, spin
polarization decay has often been assumed to be proportional to
$e^{-t/\tau_s}$, where $\tau_s$ is referred to as the \emph{spin
relaxation time}. However, if the magnitude of the spin-orbit
interaction is anisotropic (as is usually the case in systems
studied experimentally), spin-polarization decay can occur even in
the absence of momentum scattering. This process is characterized by
a non-exponential decay and is sensitive to the initial conditions,
and cannot therefore be described by a spin relaxation time. Weak
momentum scattering introduces a spin relaxation time $\tau_s
\propto \tau_p$ (unlike strong momentum scattering, which gives the
well-known~\cite{dya72, pik84} trend $\tau_s \propto \tau_p^{-1}$),
yet even in the presence of weak momentum scattering a fraction of
the polarization may survive at long times. It will emerge from our
work that, in the ballistic and weak momentum scattering regimes,
the concept of a spin relaxation time is of very limited
applicability and in general does not provide an accurate
description of the physics of spin polarization decay.

We emphasize that the results presented in this paper are true for
(delocalized) electron spins in \emph{any} nonmagnetic solid where
spin-orbit coupling is important. Since in today's experiments
mobilities range over many orders of magnitude, the results
presented are directly relevant to ongoing state-of-the-art
research.

The outline of this article is as follows. In section II we discuss
the time evolution of the density matrix, deriving an equation which
describes the return to equilibrium of a spin polarization. We
demonstrate that in the general case there exists a fraction of the
spin polarization which does not precess, and explain its relevance
to the subsequent time evolution of the spin polarization. Section
III is devoted to spin-1/2 electron systems, in which first the
known D'yakonov-Perel' limit is discussed, then the complex
situations in the ballistic and weak momentum scattering regimes are
presented. We stress the importance of non-exponential decay and of
incomplete spin dephasing. Finally, in the last part we demonstrate
that, although spin precession is qualitatively different in
spin-1/2 electron and spin-3/2 hole systems, spin polarization decay
in these systems can be understood based on the same fundamental
concepts.

\section{Time evolution of the density matrix}

We assume a nonequilibrium spin polarization has been generated in a
homogeneous, unstructured system and study its time evolution in the
absence of external fields. The system is described by a density
matrix, which in principle has matrix elements diagonal and
off-diagonal in momentum space. Since the spin operator is diagonal
in the wave vector $\bm{k}$, we will only be concerned with the
part of the density matrix diagonal in momentum space, which is
denoted by $\rho$. Henceforth, by ``density matrix" we understand
the part of the density matrix diagonal in wave vector.

The spin density is given by $\bkt{\bm{S}} \equiv \trace \bm{S} \rho
= \trace \bm{S} \bar{\rho}$, where $\bm{S}$ is the spin operator,
and the overline represents averaging over directions in momentum
space. Only the isotropic part $\bar{\rho}$ of the density matrix is
responsible for spin population decay. \cite{pik84} It is therefore
convenient to divide $\rho$ into $\rho = \bar{\rho} + g$, where $g$
is the anisotropic part of $\rho$. Based on the quantum Liouville
equation, we obtain an equation describing the time evolution of
$\rho$ (Ref.\ \onlinecite{ave02}), which in turn is split into a set
of equations for $\bar{\rho}$ and $g$ similar to those found by
Pikus and Titkov: \cite{pik84}
\begin{subequations}
\begin{eqnarray}\label{eq:rhobarga}
\pd{\bar{\rho}}{t} + \frac{i}{\hbar} \, \overline{[H, g]} & = &
0,
\\ [1ex]\label{eq:rhobargb}
\pd{g}{t} + \frac{i}{\hbar} \, [H, g] + \frac{g}{\tau_p} & = & -
\pd{\bar{\rho}}{t} - \frac{i}{\hbar} \, [H, \bar{\rho}] .
\end{eqnarray}
\end{subequations}
These equations hold both for spin-1/2 electrons and for spin-3/2
holes. We assume elastic scattering by short-range impurities,
implying that the collision term involving $\bar{\rho}$ vanishes
\cite{pik84} and the remainder is proportional to the inverse of the
scalar momentum relaxation time~\cite{multi_tau} $1/\tau_p$.

Before proceeding, we would like to make two remarks concerning the
form of the scattering term. Firstly, in the presence of spin-orbit
coupling both intraband and interband transitions exist, while we
have assumed a simplified form of the scattering term. In the
version of the relaxation time approximation employed in this work
the spin splitting of the bands is not taken into account in the
scattering term. This approximation is justified by the fact that
spin eigenstates are generally \emph{not} energy eigenstates, and it
can be straightforwardly shown, based on the theory we present, that
accounting explicitly for interband transitions will not change the
fundamental physics of spin polarization decay, rather it will only
give less transparent solutions. Furthermore, spin-flip scattering
in nonmagnetic systems is third-order in the scattering potential
and/or first order in the ratio of the spin-orbit splitting and the
kinetic energy.

Secondly, it should be noted that, for degenerate carriers, the
return to equilibrium requires energy dissipation. However, as noted
above, in a nonmagnetic material with spin-orbit coupling the spin
eigenstates characterizing the nonzero spin polarization are not
energy eigenstates. On the other hand, unlike in, e.g., nuclear
systems, the nonthermal energy characterizing this nonequilibrium
configuration is essentially a kinetic energy, but it is not in the
spin degree of freedom. Therefore, energy dissipation has no
qualitative effect for the main conclusions in our paper.

A solution to Eq.\ (\ref{eq:rhobargb}) can be obtained by making the
transformation $g = e^{- i H t/\hbar} g_I \, e^{i H t/\hbar}$, which
is analogous to the customary switch to the interaction picture.
This transformation turns Eq.\ (\ref{eq:rhobargb}) into an equation
for $g_I$
\begin{equation}
\pd{g_I}{t} + \frac{g_I}{\tau_p} = - \pd{\bar{\rho}_I}{t},
\end{equation}
where $\bar{\rho}_I$ is defined by $\bar{\rho} = e^{- i H t/\hbar}
\bar{\rho}_I \, e^{i H t/\hbar}$. Treating the RHS as a source
term, this equation allows an analytical solution using an
integrating factor. Substituting this solution into Eq.\
(\ref{eq:rhobarga}) yields
\begin{equation}\label{eq:rhobarJ}
\arraycolsep 0.1ex
\begin{array}{rl}
\displaystyle \pd{\bar{\rho}}{t} + \frac{i}{\hbar\tau_p}
\int_0^{t} \! dt' & \displaystyle e^{- (t - t')/\tau_p} \,
\overline{e^{- i H (t - t')/ \hbar} \, [H, \bar{\rho}(t') ] \,
e^{i H (t - t') / \hbar}}
\\ [2.2ex] & \displaystyle
= -  \frac{i}{\hbar} \, e^{-t/\tau_p}\,
\overline{e^{-i H t / \hbar} \, [H, \rho_0]\, e^{i H t / \hbar}} ,
\end{array}
\end{equation}
where $\rho_0$ is the initial value $\rho (t=0)$. This equation is
the main result of our paper. It describes the precession-induced
decay of spin polarization in all regimes of momentum scattering for
any nonmagnetic solid state system with spin-orbit interactions.
This equation does not anticipate any particular form of spin
polarization decay, such as exponential decay.

The form of the initial density matrix $\rho_0$ is important and
lies at the root of the novel physics discussed in this paper. In
general $\rho_0$ has two contributions, $\rho_0 = \rho_{0\|} +
\rho_{0\perp}$. The component $\rho_{0\|}$ commutes with $H$ and is given
by $\rho_{0\|} = ({\rm tr}\, \rho_0 H / {\rm tr} \, H^2)\, H$, in a
generalization of Gram-Schmidt orthogonalization. $\rho_{0\perp}$ is
simply the remainder, and it satisfies the condition ${\rm tr}\,
\rho_{0\perp} H = 0$. $\rho_{0\|}$ is a matrix that is parallel to the
Hamiltonian, and represents the fraction of the initial spin
polarization that does not precess, or alternatively the fraction of
the initial spins that are in eigenstates of the Hamiltonian.
$\rho_{0\perp}$ is orthogonal to the Hamiltonian, and represents the
fraction of the initial spin polarization that does precess.

\section{Spin-1/2 electron systems}

First we discuss Eq.\ (\ref{eq:rhobarJ}) for spin-1/2 systems. The
Hamiltonian describing spin-orbit coupling has the form $H =
(\hbar/2) \, \bm{\sigma}\cdot\bm{\Omega} (\bm{k})$ and $\bar{\rho}$
may be decomposed as $\bar{\rho} = \frac{1}{2} \, [n +
\bm{s}(t)\cdot \bm{\sigma}]$, where $n$ represents the number
density and $\bm{s}(t)$ the spin polarization. Equation
(\ref{eq:rhobarJ}) has qualitatively different solutions depending
on the regime under study, and they are discussed in detail below.

\subsection{Exponential decay in the strong momentum scattering regime}

A solution to Eq.\ (\ref{eq:rhobarJ}) characterizing
\emph{relaxation} is understood as \emph{exponential} decay of the
form $\bar{\rho} (t) = e^{-\Gamma_s t}\,\bar{\rho}_0 $, where $\Gamma_s$
is generally a second-rank tensor that represents the inverse of the
spin relaxation time $\tau_s$. Such a simple solution of Eq.\
(\ref{eq:rhobarJ}) does \emph{not} exist in general, but for strong
momentum scattering ($\Omega \tau_p \ll 1$) the RHS of Eq.\
(\ref{eq:rhobarJ}) can be neglected. Then substituting for
$\bar{\rho}$ and $H$ in Eq.\ (\ref{eq:rhobarJ}) yields the DP
expression \cite{dya72, pik84} for $\Gamma_s$, which may be written
as $(\Gamma_s)_{ij} = \tau_p \, \big( \overline{\Omega^2\delta_{ij}
- \Omega_i \Omega_j} \big)$, where $i,j=x,y,z$. Strong momentum
scattering yields exponential spin relaxation and the
well-known~\cite{dya72, pik84} trend $\tau_s \propto \tau_p^{-1}$ .

\subsection{Oscillations in the ballistic regime}

Previously, most analytical studies have focused on strong momentum
scattering. \cite{dya72, dya86, pik84, ave99, ave02, ave06, qi03,
lau01, kai04} We will show that the ballistic and weak momentum
scattering regimes are far more complex. \cite{grim05, schw06,
gri01, gla07} In the ballistic limit $\tau_p \rightarrow \infty$ and
Eq.\ (\ref{eq:rhobarJ}) can be solved exactly as $\bar{\rho} (t) =
\overline{e^{-i H t'/\hbar} \, \rho_{0\perp} \, e^{i H t'/\hbar}} +
\bar{\rho}_{0\|}$, which can also be obtained from the quantum
Liouville equation. \cite{dim06} This determines the time evolution
of an initial spin polarization $\bm{s} (t=0) = \bm{s}_0$, i.e., the
component of $\bm{s} (t)$ along $\bm{s}_0$. For simplicity
$\bm{s}_0$ is here assumed independent of $\bm{k}$; a
$\bm{k}$-dependent distribution would not change the results
qualitatively. From the solution for $\bar{\rho}$ in the ballistic
limit we have
\begin{equation}\label{eq:spinintime}
\bm{s}(t) \cdot \hat{\bm{s}}_0 = \overline{[1 - (\hat{\bm{\Omega}}
\cdot \hat{\bm{s}}_0)^2]\,\cos \Omega t} +
\overline{(\hat{\bm{\Omega}} \cdot \hat{\bm{s}}_0)^2},
\end{equation}
where $\hat{\bm{a}}$ denotes the unit vector in the direction of
$\bm{a}$. The last term corresponds to $\bar{\rho}_{0\|}$. It is best
to take a concrete example, such as the Hamiltonian of a 2D system
on a (001) surface with linear Rashba \cite{ras84} and Dresselhaus
\cite{dre55} spin-orbit interactions, $H = \alpha (\sigma_x k_y -
\sigma_y k_x) + \beta (\sigma_x k_x - \sigma_y k_y) $. We consider
first effective fields $\bm{\Omega} (\bm{k})$ such that the
magnitude $|\bm{\Omega} (\bm{k})|$ does not depend on the direction
of $\bm{k}$, for example either $\alpha = 0$ or $\beta = 0$ yields
$|\bm{\Omega} (\bm{k})| = \Omega (k)$. In this case the initial spin
polarization will simply oscillate with frequency $\Omega$. It is
helpful to visualize a population of spins on the Fermi surface, all
initially pointing up. If $|\bm{\Omega} (\bm{k})|$ is the same at
all points $\bm{k}$ on the Fermi surface, all spins that were in
phase initially will be in phase again after one precession period.
Some fraction of the initially oriented spins $\bm{s}_0$,
corresponding to the last term in Eq.\ (\ref{eq:spinintime}), has a
nonzero overlap with the local field $\bm{\Omega} (\bm{k})$ so that
the projection of $\bm{s}_0$ on $\bm{\Omega} (\bm{k})$ will be
preserved. This fraction is zero if the initial spin is out of the
plane, but significant if it is in the plane.

\subsection{Non-exponential decay in the ballistic regime}

The case when $|\bm{\Omega} (\bm{k})|$ depends on the direction of
$\bm{k}$ is of great relevance to experiment, where spin-orbit
coupling is rarely attributable to a single mechanism. Spins on the
Fermi surface precess with incommensurable frequencies and once they
are out of phase they never all get in phase again (but the
polarization fraction due to $\rho_{0\|}$ is conserved.) In our
example, if $\beta \ll \alpha$ we can write $\Omega \approx
\bar{\Omega} \, (1 + \eta \sin 2\phi)$, where $\bar{\Omega} \equiv k
\, \sqrt{\alpha^2 + \beta^2}$, the small parameter $\eta \equiv
\alpha\beta / [2 (\alpha^2 + \beta^2)]$, and $\phi$ is the polar
angle of $\bm{k}$. The motion of the spins, given by Eq.\
(\ref{eq:spinintime}), is averaged over the Fermi circle. Consider
the term $\overline{\cos \Omega t}$ in Eq.\ (\ref{eq:spinintime}) as
an example. The angular average yields $\cos (\bar{\Omega} t) \, J_0
(\eta \bar{\Omega} t)$, where $J_0$ is a Bessel function of the
first kind. This function has the form of a decaying oscillation but
it does \emph{not} reduce to an exponentially damped oscillation in
any limit. At long times we have $J_0 (\eta \bar{\Omega} t)
\rightarrow \sqrt{2/(\pi \eta\bar{\Omega} t)}\, \cos
(\eta\bar{\Omega} t - \pi/4)$. A similar, more complicated
expression in terms of Bessel functions applies for the remaining
term $\overline{( \hat{\bm{\Omega}} \cdot \hat{\bm{s}}_0)^2 \,\cos
\Omega t}$ in Eq.\ (\ref{eq:spinintime}).

The anisotropy of the Fermi surface introduces a mechanism for
non-exponential spin decay \cite{gla05} with a characteristic time
$\tau_d \propto (\eta \bar{\Omega})^{-1}$, referred to as the
dephasing time $\tau_d$. For pure spin dephasing, i.e., in the
absence of momentum scattering, two limiting cases can be
distinguished. If $\bm{\Omega} (\bm{k}) \perp \bm{s}_0$ for all
$\bm{k}$ (e.g., a spin orientation perpendicular to the 2D plane on
a [001] surface), spin dephasing reduces the spin polarization to
zero. On the other hand, spin dephasing is completely suppressed if
$\bm{\Omega} (\bm{k}) \parallel \bm{s}_0$ for all $\bm{k}$ (e.g., a
2D electron system in a symmetric quantum well on a [110] surface
with a spin orientation perpendicular to the 2D plane \cite{dya86}).
In general (in particular for 3D systems), an intermediate situation
is realized where the spin polarization is reduced because of
dephasing, but it remains finite. The surviving part is identified
with $\rho_{0\|}$ in the initial density matrix. This process is
referred to as \emph{incomplete} spin dephasing.

\begin{figure}[tbp]
 \includegraphics[width=1.0\columnwidth]{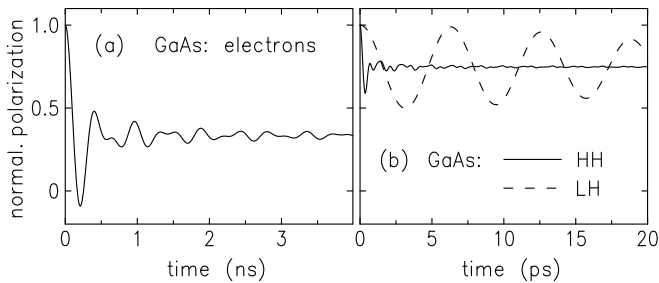}
 \caption{\label{fig:deph} Incomplete spin dephasing of (a) electron
 spins in the $k^3$-Dresselhaus model and (b) of heavy-hole spins
 (solid lines) and light-hole spins (dashed lines) in bulk GaAs in
 the ballistic limit. The vertical axis shows the normalized spin
 polarization $\bm{s}(t) \cdot \hat{\bm{s}}_0 / |\bm{s}_0|$. The
 initial spin polarization $\bm{s}_0$ is assumed to point along
 [001]. The Fermi wave vector is $k_F = 10^8$m$^{-1}$. Note the
 different time scales in (a) and (b).}
\end{figure}

Analogous results hold for the $k^3$-Dresselhaus model,
\cite{dre55} but the terms leading to dephasing cannot be
expressed in a simple form due to the complex angular dependence
of $|\bm{\Omega} (\bm{k})|$ on the direction of $\bm{k}$.
Figure~\ref{fig:deph}(a) shows the incomplete dephasing of
electron spins in bulk GaAs calculated using the $k^3$-Dresselhaus
model. At long times the initial spin polarization settles to a
value $\approx 0.33$, which is independent of any system
parameters, including the spin-orbit constant.

\subsection{Weak momentum scattering regime}

In the regime of weak momentum scattering the solution to Eq.\
(\ref{eq:rhobarJ}) may be written approximately as
\begin{equation}\label{eq:rhobarweak}
\bar{\rho} (t) = \bar{\rho}_{0\|} + e^{-t/\tau_p} \, \overline{e^{-i
H t'/\hbar} \, \rho_{0\perp} \, e^{i H t'/\hbar}}.
\end{equation}
Since the momentum scattering rate $1/\tau_p$ is small, the term
under the overline is taken to lowest order in $1/\tau_p$. The
second term on the RHS of Eq.\ (\ref{eq:rhobarweak}) describes
damped oscillations with amplitude decaying exponentially on a scale
$\propto \tau_p$. This trend is the \emph{inverse} of that for
strong momentum scattering and is explained by the following
argument. If one spin, precessing on the Fermi surface in phase with
all the other spins, is scattered to a different wave vector, it
will precess about a different effective field and will no longer be
in phase with the other spins. Thus the combined effect of spin
precession and momentum scattering---even when the latter is only
weak---reduces the spin polarization faster. The fraction of the
spin polarization corresponding to $\rho_{0\|}$ survives. This
remaining polarization decays via spin-flip scattering (the
Elliott-Yafet mechanism~\cite{ell54, yaf63}) on much longer time
scales.

An exception occurs when $\bm{\Omega} (\bm{k}) \parallel \bm{s}_0$
for \emph{all} spins. This situation is realized, e.g., for a 2D
electron system in a symmetric quantum well on a [110] surface with
a spin orientation perpendicular to the 2D plane. For this
particular case it is well known that spin precession and momentum
scattering do not affect at all the initial spin
orientation. \cite{dya86, ohn99} From the preceding discussion we
can understand this by noting that the initial density matrix
$\rho_0 = \rho_{0\|}$ commutes with the spin-orbit Hamiltonian and
Eq.\ (\ref{eq:rhobarJ}) shows that $\bar{\rho} = \bar{\rho}_{0\|}$ for
all times. The polarization decays eventually via spin-flip
scattering. \cite{ell54, yaf63}

In the weak momentum scattering regime for non-negligible anisotropy
the spin decay rate is determined by the larger of $\eta\,
\bar{\Omega}$ and $1/\tau_p$. Momentum scattering introduces an
exponential envelope but in this limit the concept of a spin
relaxation time is evidently of limited use. \cite{linprop} In
spin-1/2 systems dephasing will be important for high-mobility
carriers. Results consistent with our findings were obtained
experimentally by Brand \textit{et al.}\ \cite{bra02} who studied
the oscillatory time evolution of an optically-generated spin
polarization in a high-mobility 2D electron system in a GaAs/AlGaAs
quantum well. Similarly, in materials in which a nonequilibrium spin
density is excited, the time evolution of this spin density can be
studied, for example, by means of magnetic circular dichroism
techniques. \cite{sra06}

We have assumed an initial spin distribution sharp at the Fermi
edge. In practice this distribution spans a window in
$\bm{k}$-space, introducing additional dephasing between spins at
wave vectors of slightly different magnitudes. In this case even
an isotropic spin splitting leads to decay, though the
polarization due to $\rho_{0\|}$ is still robust. For example, in
2D for isotropic spin splitting, $\int dk \, k \, \cos \Omega t
\propto t^{-2}$, so that the spin polarization, instead of
oscillating indefinitely, decays as $t^{-2}$.

We have also assumed the initial spin distribution to be independent
of wave vector. The theory is well-equipped to deal with wave-vector
dependent spin polarizations. (Indeed, the initial spin distribution,
contained in the density matrix $\rho_0$, is in general wave
vector-dependent.) The wave vector-independent cases discussed at
length are intended as examples, and they have been selected as more
straightforward cases for clarity.

\section{Spin-3/2 hole systems}

Next we discuss spin-3/2 hole systems, which are different from
spin-1/2 electron systems for several reasons. The presence of extra
terms in the spin density matrix of spin-3/2 systems (in addition to
the number density and spin polarization) has important consequences
for spin dynamics. \cite{win04b, dim06} Spin-orbit coupling affects
the energy spectrum in the valence band to a greater extent, and the
spin orientation often disappears on scales comparable to $\tau_p$.
The relation $\Omega \, \tau_p \ll 1 $ holds less frequently than
for electrons in systems accessible experimentally.

We consider spin-3/2 holes to be described by the Luttinger
Hamiltonian, \cite{lut56, bal74}
\begin{equation}\label{eq:Lutt}
H_0 = \frac{\hbar^2}{2m_0} \left[\left (\gamma_1 +
{\textstyle\frac{5}{2}} \bar{\gamma} \right)k^2 - 2\bar{\gamma}
(\bm{k} \cdot \bm{S})^2 \right] - H_C,
\end{equation}
where $m_0$ is the bare electron mass, $\bm{S}$ the spin operator
for effective spin 3/2, $\bar{\gamma} \equiv (2\gamma_2 + 3\gamma_3)
/ 5$, and $\gamma_1$, $\gamma_2$, and $\gamma_3$ are Luttinger
parameters. $H_C$ represents the anisotropic terms with cubic
symmetry \cite{bal74} which will be given below.

We work first in the spherical approximation in which $H_C$ is
neglected. The energy dispersions $E_\mathrm{HH}$ for the heavy
holes (HHs, spin projection in the direction of $\bm{k}$ is $m_s =
\pm 3/2$) and $E_\mathrm{LH}$ for the light holes (LHs, $m_s = \pm
1/2$) are $E_\mathrm{LH/HH} (\bm{k}) = \hbar^2k^2\, (\gamma_1 \pm
2\bar{\gamma}) /({2m_0})$.

\subsection{Exponential decay in the strong momentum scattering regime}

In the strong momentum scattering limit an exponential solution
$\bar{\rho} (t) = e^{-\Gamma_s t}\,\bar{\rho}_0 $ is possible. The
tensor $\Gamma_s = \tau_s^{-1} \, \openone$, showing that the
relaxation times for all spin components are equal,
\begin{equation}
\frac{1}{\tau_s} = {\textstyle\frac{2}{5}} \, \Omega^2\, \tau_p
= \frac{8}{5} \left(\frac{\hbar \bar{\gamma} k^2}{m_0}\right)^2 \tau_p,
\end{equation}
where now the frequency $\Omega (\bm{k}) = (E_\mathrm{LH} -
E_\mathrm{HH}) / \hbar = 2\hbar\bar{\gamma} k^2/m_0$ corresponds to
the energy difference between the HH and LH bands. \cite{dim06}
Despite the qualitatively different spin precession, the situation
is rather similar to electron spin relaxation and can be explained
in terms of the same random walk picture familiar from the study of
electron spin relaxation.

\subsection{Ballistic and weak momentum scattering regimes in the spherical approximation}

In the ballistic limit Eq.\ (\ref{eq:rhobarJ}) is again solved by
$\bar{\rho} (t) = \overline{e^{-i H t'/\hbar} \, \rho_0 \, e^{i H
t'/\hbar}}$. An initial spin polarization will oscillate
indefinitely since $\Omega$ is the same for all holes on the Fermi
surface. For weak momentum scattering Eq.\ (\ref{eq:rhobarweak})
applies to holes also. The spin polarization consists of damped
oscillations, decaying on a time scale $\propto \tau_p$, plus a term
corresponding to $\rho_{0\|}$, which survives at long times.
$\rho_{0\|}$ does not depend on the Luttinger parameters or the Fermi
wave vector and will therefore be the same in any system described
by the Luttinger Hamiltonian. This remaining polarization decays via
spin-flip scattering as discussed in Refs.~\onlinecite{uen90, fer91}.

\subsection{Cubic-symmetry terms and dephasing}

Dephasing is introduced if the term $H_C$ with cubic symmetry
\cite{bal74} is included in the Luttinger Hamiltonian,
\begin{equation}\label{eq:HC}
H_C = \frac{\hbar^2 \Delta}{m_0} \left( k_x k_y \{J_x, J_y \} +
k_y k_z \{J_y, J_z \} + k_z k_x \{J_z, J_x \} \right),
\end{equation}
where $\Delta = (\gamma_3 - \gamma_2)/2$. The eigenenergies are now
given by $E_\mathrm{HH/LH} = \gamma_1\hbar^2k^2/ 2m_0 \pm
2\sqrt{E^2_\mathrm{an}}$, where
\begin{equation}
E_\mathrm{an}^2 = \bar{\gamma} k^4 - 2\bar{\gamma}\Delta
\left[k^4 - 6 \left(k_x^2k_y^2 + k_y^2k_z^2 + k_z^2k_x^2\right)\right]
+ k^4 \Delta^2.
\end{equation}
The cubic-symmetry terms contained in Eq.\ (\ref{eq:HC}) are usually
neglected in charge and spin transport without a significant loss of
accuracy. However, they play a crucial role in spin relaxation in
the weak momentum scattering regime, which for holes extends over a
wide range of $k$.

Due to the presence of $H_C$, the energy dispersion relations and
therefore $|\bm{\Omega} (\bm{k})|$ depend on the direction of
$\bm{k}$, causing an initial spin polarization to decay even in
the ballistic limit, where incomplete spin dephasing occurs. Our
numerical calculations exemplified in Fig.~\ref{fig:deph}(b) show
that an initial spin polarization falls to a fraction much higher
than in the electron cases studied. It decays more slowly for the
LHs, for which the Fermi surface is nearly spherical, than for the
HHs, for which the Fermi surface deviates significantly from a
sphere. At long times the initial spin polarization settles to a
value $\approx 0.75$, which is independent of any system
parameters, including the Luttinger parameters.

\section{Summary}

In conclusion, we have shown that the decay of spin polarization in
semiconductors, brought about by the interplay of spin precession
and momentum scattering, depends strongly on the regime of momentum
scattering. In the ballistic regime the spin polarization decays via
a dephasing mechanism which is present due to the fact that the
magnitude of the spin-orbit interaction generally depends on the
direction of the wave vector. This mechanism may reduce a spin
polarization to zero (complete dephasing) or a fraction of the
initial value (incomplete dephasing). Weak momentum scattering
destroys an initial spin polarization, whereas strong momentum
scattering helps maintain an initial polarization.

The authors are grateful to S.~Lyon for a critical reading of the
manuscript and to E.~Rashba, A.~H.~MacDonald, Q.~Niu, G.~Vignale,
J.~Sinova, B.~A.~Bernevig, C.~H.~Chang, and A.~Mal'shukov for
enlightening discussions. The research at Argonne National
Laboratory was supported by the U.S. Department of Energy, Office of
Science, Office of Basic Energy Sciences, under Contract No.\
DE-AC02-06CH11357. Research at Northern Illinois University was
supported by the Department of Education.


\begin{thebibliography}{10}
\bibitem{multi_tau} In a more elaborate approach different
 relaxation times characterize the spherical components of $\rho$
 and corrections due to spin-orbit coupling are present in the
 collision term. It can be shown that neither change affects the
 physics qualitatively.
\bibitem{linprop} The relation $\tau_s \propto \tau_p$ was also
 found for the specific case of the Rashba Hamiltonian in Ref.\
 \onlinecite{grim05}, and more generally in Ref.\
 \onlinecite{schw06}, although neither paper discusses oscillations
 in the spin polarization. See also Refs.\ \onlinecite{gri01,
 gla07}.
\end{thebibliography}
\end{document}